\documentclass[%
reprint,
unsortedaddress,
amsmath,amssymb,
prb,
]{revtex4-1}

\usepackage{graphicx}
\usepackage[colorlinks, bookmarks=false,linkcolor=blue, urlcolor=blue, plainpages=false, pdfpagelabels]{hyperref}
\usepackage[version=3]{mhchem}
\usepackage{wasysym}
\usepackage{youngtab}
\usepackage{young}

\renewcommand*{\H}{\hat{H}}
\renewcommand*{\P}{\hat{P}}
\renewcommand*{\S}{\hat{S}}

\newcommand*{\irb}[1]{\left(#1\right)}
\newcommand*{\dsum}[1]{\displaystyle\sum_{#1}}
\renewcommand*{\exp}[1]{\,\mathrm{e}^{#1}}

\renewcommand*{\c}[1]{\hat{c}_{#1}^{\vphantom{\dagger}}}
\newcommand*{\cd}[1]{\hat{c}_{#1}^{\dagger}}
\newcommand*{\SU}[2][{}]{\ensuremath{\mathrm{SU}\hspace{-0.5mm}\irb{#2}_{#1}}}

\newcommand*{\zzz}{000}
\newcommand*{\pzz}{\pi 00}
\newcommand*{\zpp}{0\pi\pi}
\newcommand*{\ppp}{\pi\pi\pi}

\newcommand*{\eref}[1]{Eq.\ref{eq:#1}}
\newcommand*{\gref}[1]{FIG.~\ref{fig:#1}}
\newcommand*{\citer}[1]{Ref.[\onlinecite{#1}]}

\newcommand*{\ie}{\textit{i.e.}\ }

\begin{document}

\title{Stabilization of the chiral phase of the \SU{6m} Heisenberg model on the honeycomb lattice with $m$ particles per site for $m$ larger than $1$}

\author{J\'er\^ome Dufour}
\author{Fr\'ed\'eric Mila}
\affiliation{Institute of Physics, \'Ecole Polytechnique F\'ed\'erale de Lausanne (EPFL), CH-1015 Lausanne, Switzerland }

\date{\today}
\begin{abstract}
We show that, when $N$ is a multiple of 6 ($N=6m$, $m$ integer), the \SU{N}
Heisenberg model on the honeycomb lattice with $m$ particles per site has a
clear tendency toward chiral order as soon as $m\geq 2$. This conclusion has
been reached by a systematic variational Monte Carlo investigation of
Gutzwiller projected wave-functions as a function of $m$ between the case of
one particle per site ($m=1$), for which the ground state has recently been
shown to be in a plaquette singlet state, and the $m\rightarrow \infty$ limit,
where a mean-field approach has established that the ground state has chiral
order. This demonstrates that the chiral phase can indeed be stabilized for not
too large values of $m$, opening the way to its experimental realisations in
other lattices.
\end{abstract}

\maketitle

\section{Introduction}
Progress in cold atoms experiments has opened the door to new and exciting
physics~\cite{Sugawa2010, Taie2012, Pagano2014, Cazalilla2014, Zhang2014,
Scazza2014, Hofrichter2016}. When fermionic ultracold alkaline-earth atoms with
nuclear spin $I$ are trapped in optical lattices, the physics is governed by a
generalized Hubbard model with $N=2I+1$ colors (or flavors) of fermionic
particles~\cite{Wu2003, Honerkamp2004, Cazalilla2009, Gorshkov2010}. In the
limit of strong on-site repulsion, when the optical wells are deep, and if the
number of atoms per site is an integer, this system is in a Mott insulating
phase and the low energy physics is effectively described by the \SU{N}
Heisenberg model:
\begin{equation}\label{eq:Heisenberg-general}
	\H = \dsum{<i,j>}\dsum{\alpha\beta}\S^{\alpha\beta}_{i}\S^{\beta\alpha}_{j}.
\end{equation}
This Hamiltonian is known to exhibit various ground states depending not only
on the lattice but also on the number of colors $N$ and on the on-site \SU{N}
symmetry of the wave function, \ie the irreducible representation (irrep)
labelled by a Young tableau with $m$ boxes. For ultra-cold fermions, the case
of $m$ atoms per well corresponds to the fully antisymmetric irrep with a
Young tableau consisting of a single column with $m$ boxes. A non exhaustive
list of exotic ground states contains $N$-flavor liquids with algebraic
correlations~\cite{Marston1988, Assaad2005, Xu2010, Corboz2012a, Cai2013},
different N\'eel-type states with long range order~\cite{Toth2010, Corboz2011,
Bauer2012a}, translational symmetry breaking states like generalized
valence-bond solids~\cite{Corboz2007, Arovas2008, Hermele2009, Hermele2011,
Corboz2012, Song2013, Corboz2013a} or $d$-merisation ($d=2,3$) on a
chain~\cite{Dufour2015} and chiral spin liquids~\cite{Wen1989, Hermele2009,
Hermele2011, Szirmai2011, Bieri2012, Cai2013, Song2013}.

The possibility to stabilize chiral phases is particularly interesting since
the search for experimental realisations in lattice models is still on-going.
However, for the simple \SU{N} model with only nearest-neighbor Heisenberg
interactions on the square lattice, unambiguous evidence of chiral order has
only been obtained in the limit where $N$ and $m$ tend to infinity keeping the
ratio $k\equiv N/m$ fixed when $k>4$\cite{Hermele2009}. For $m=1$, other types
of order can be stabilized. For instance, exact diagonalisation results
suggest that color order might be stabilized for \SU{5}, while spontaneous
dimerisation seems to take place for \SU{8}. These results clearly call for
additional investigations of the properties of the model as a function of $m$
for fixed $k$.

In this paper, our aim is to address this issue in the context of the \SU{N}
Heisenberg models on the honeycomb lattice. This model has been studied by
\citet{Szirmai2011, Sinkovicz2013} for $k=6$ in the mean-field limit where $N$
and $m$ tend to infinity keeping the ratio $k$ fixed. They found a chiral
ground state with $2\pi/3$ flux per hexagonal plaquette, and two plaquette
states with higher energies: the $\zpp$ plaquette in which each $0$ flux
plaquette is surrounded by $\pi$ flux plaquettes, and a $\zzz$ plaquette phase
in which hexagons are completely decoupled. However, a recent
work~\cite{Nataf2016} done on the same lattice for \SU{6} in the fundamental
irrep has given strong evidence in favor of the $\zpp$ plaquette ground state
over the chiral one with $2\pi/3$ flux. This irrep corresponds to a physical
model with $m=1$ particle per site for which \eref{Heisenberg-general} reduces
to a simple permutation Hamiltonian: $\H_{\P} = \sum_{<i,j>}\P_{ij}$, where
the constant has been omitted. It was also shown in \citer{Nataf2016} that the
chiral phase can be stabilized by adding a ring-exchange term $\H_{\hexagon} =
i\sum_{\hexagon} (\P_{\hexagon}^{\vphantom{1}} - \P_{\hexagon}^{-1})$ to the
Hamiltonian. Note that plaquette and chiral states can easily be distinguished
experimentally by their very different signatures in the spin structure factor,
as discussed in Refs.[\onlinecite{Szirmai2011}] and
[\onlinecite{Sinkovicz2013}].

Here we explore another path to the possible stabilisation of a chiral phase
for the \SU{N} Heisenberg model. Instead of adding a ring-exchange term, we
increase the number of particles per site $m$ for $k=6$, and we study fully
antisymmetric irreps on each site labelled by a Young tableau with $m$ boxes
in one column: 
\begin{align*}
	\yng(1) &\; \SU{6} &
	\yng(1,1) &\; \SU{12} &
	\yng(1,1,1) &\; \SU{18} ... 
\end{align*}
For this family of irreps, the fermionic operators $\cd{},\c{}$ together with
the identity:
\begin{equation}\label{eq:Sab=aaab-m/Ndab}
	\S^{\alpha\beta} = \cd{\alpha}\c{\beta} - \dfrac{m}{N}\delta_{\alpha\beta}
\end{equation}
allow the rewriting of \eref{Heisenberg-general} as:
\begin{equation} \label{eq:Heisenberg-fermionic-operator}
	\H = \dsum{<i,j>}\dsum{\alpha\beta}\cd{i\alpha}\c{i\beta}\cd{j\beta}\c{j\alpha}
\end{equation}
where the constant $-zm^{2}/2N$ has been dropped. In the following, our goal
is to investigate how the system evolves between the chiral ground state of
\citer{Szirmai2011} for $m=\infty$ and the $\zpp$ plaquette ground state of
\citer{Nataf2016} obtained for $m=1$. We will present numerical results that
give strong indication that the chiral phase is stabilized for $m>1$.

\section{VMC results}
\subsection{Method}
Since our aim is to systematically study the fully anti-symmetric irreps of the
\SU{6m} Heisenberg model on the honeycomb lattice, we need a numerical method
that works for any $m$. Quantum Monte Carlo suffers from the sign problem, ED
does not give access to large enough clusters when $m>1$, and iPEPS has so far
only given results for $m=1$. The variational Monte Carlo (VMC)
method~\cite{Yokoyama1987, Gros1989} is therefore the only reliable numerical
method that was proven to be efficient to study more complicated
representations~\cite{Paramekanti2007} like the fully anti-symmetric
irreps~\cite{Dufour2015}. It is not limited by the system size and recovers the
mean-field results when $N$ and $m$ are large.

To have meaningful results we need to define a representative set of
variational wave functions. Following other papers~\cite{Corboz2013a,
Nataf2016} and inspired by the mean-field results~\cite{Szirmai2011}, we have
tested five different variational wave functions represented in
\gref{honeycomb-k6-wave-functions}, one chiral and four plaquette states. The
chiral wave function is the only one having no variational parameter. It has
uniform hopping amplitudes but non-uniform phase factors that creates a
homogeneous flux of $2\pi/3$ per hexagonal plaquette. This wave function does
not break the lattice symmetry but breaks the time reversal one. We want the
other wave functions to preserve the time reversal symmetry, therefore the only
allowed fluxes are $0$ and $\pi$. To preserve the rotation symmetry, and since
we chose unit cells containing at most $12$ sites, there are only $4$
non-equivalent flux configurations. Two of them have already been introduced,
$\zpp$ and $\zzz$, while the other two are: $\pzz$ consisting of a central
hexagon with $\pi$ flux surrounded by $0$ fluxes and $\ppp$ having a
homogeneous $\pi$ flux in each hexagon. Due to the simplicity of the chosen
flux configurations, there are only two meaningful variational parameters,
$t_{h}$ and $t_{d}$: the hopping terms around the central hexagons $t_{h}$, and
the hopping terms linking these hexagons $t_{d}$. Using additional hopping
terms would break other symmetries, for instance having different hopping terms
around the central hexagon would break the rotational symmetry. Since the
honeycomb lattice is bipartite, only the relative sign of $t_{d}$ and $t_{h}$
matters, therefore, for the plaquette wave functions, we have a single
variational free parameter, the ratio $t_{d}/t_{h}$ with fixed $t_{h}=-1$.

\begin{figure}[h]
	\centering
	\includegraphics[scale=0.62]{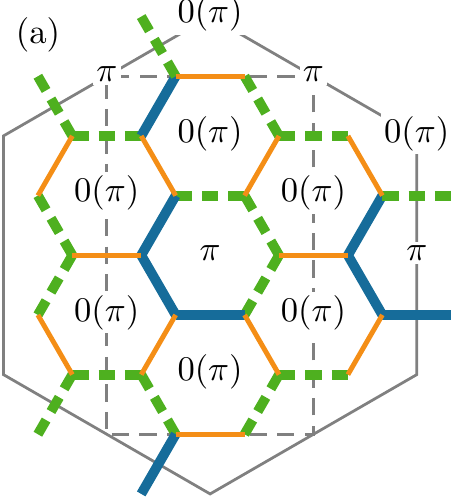}\hfill
	\includegraphics[scale=0.62]{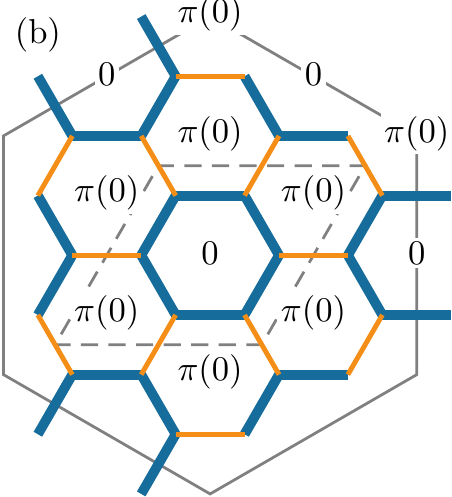}\hfill
	\includegraphics[scale=0.62]{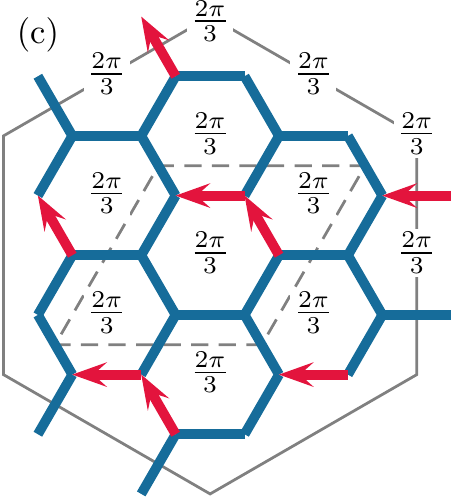}
	\caption{
	Representation of the five wave functions with their unit cell on a 24-site
	cluster with periodic boundary conditions. (a)-(b) Plaquette wave
	functions, where solid blue and dashed green bonds stand for $t_{h}$ and
	$-t_{h}$ respectively, while the thin yellow ones stand for $t_{d}$. By
	changing the sign of the ratio $t_{d}/t_{h}$ from negative to positive, the
	flux configuration of the wave function will change from $\pzz$ to $\ppp$
	in (a) and from $\zpp$ to $\zzz$ in (b). (c) Chiral wave function for which
	the hopping terms are given by $t_{ij}=t_{h}\exp{i2\pi/3}=t_{ji}^{*}$
	between the sites $i$ and $j$ connected by a red arrow and by $t_{h}$
	otherwise. The flux per plaquette is defined mod $2\pi$.
	}
	\label{fig:honeycomb-k6-wave-functions}
\end{figure}

\subsection{Results}
In this section, results on a 72-site cluster with anti-periodic boundary
conditions will first be presented. Then a finite size analysis on a
representative example, $m=3$, will show how accurately the thermodynamic
energies can be extracted. This accuracy allows us to draw conclusions on how
the VMC recovers the mean-field limit and gives new indications of a chiral
phase for $m>1$.

\gref{honeycomb-k6-n72-all-wave-functions} shows the variational energies as a
function of $m$ for different values of the ratio $t_{d}/t_{h}$ for all
different wave functions on a 72-site cluster with anti-periodic boundary
condition. This particular choice of boundary conditions allows us to measure
the energy for $t_{d}/t_{h} =-1$ because it lifts the well-known degeneracy at
the Fermi level, an important requirement to construct Gutzwiller projected
wave functions. However, it does not lift the degeneracy for the value
$t_{d}/t_{h} = 1$. The other missing energy is for $t_{d}/t_{h} = 0$ because
the VMC fails to find a well defined starting configuration, since all
hexagons are disconnected. On the lowest plot, the chiral energies are also
shown as straight lines.

\begin{figure}[h!]
	\centering
	\includegraphics[scale=0.7]{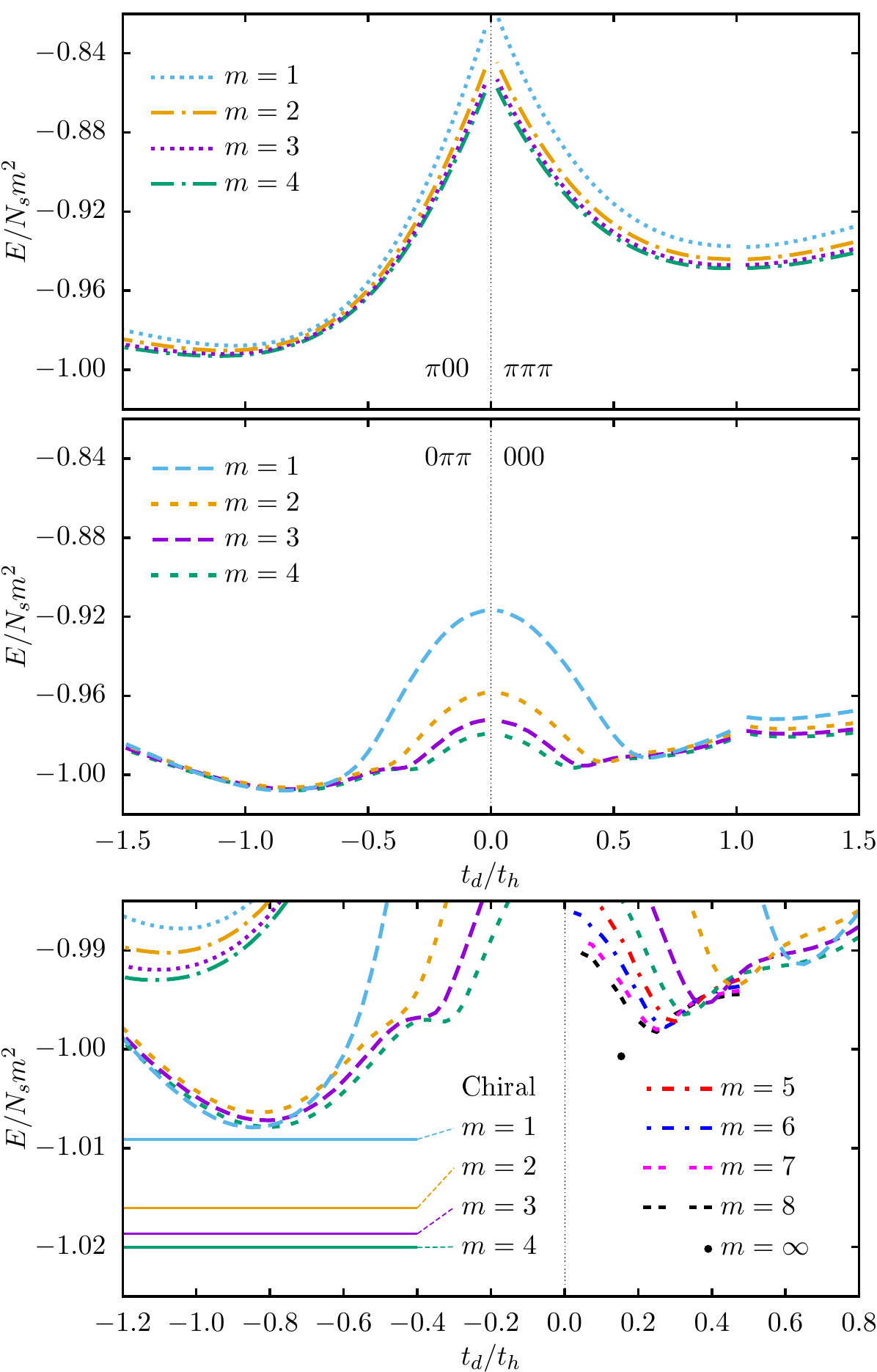}
	\caption{
	Energy per site for a cluster of $72$ sites. The upper and middle plots
	are given on the same scale to allow better comparison while the lower
	plot is a zoom around the minima. The error bars are smaller than the
	thickness of the lines. 
	Upper plot: Energy for different values of $m$ of the $\pzz$ and $\ppp$
	wave functions on the left and on the right respectively.
	Middle plot: Energy for different values of $m$ of the $\zpp$ and $\zzz$
	wave functions on the left and on the right respectively. 
	Lower plot: Energy of all the wave functions discussed in this paper. The
	energies of the chiral wave functions are represented with straight lines.
	Additional values for the $\zzz$ wave function for $5\leq m\leq8$ have been
	included. The filled circle corresponds to the minimal energy for the
	$\zzz$ wave function obtained by extrapolation in the limit $m\to\infty$.
	For larger systems, the extrapolated value of $t_d/t_h$ for the $\zzz$ wave
	function tends to zero, as it should since the mean-field result
	corresponds to isolated plaquettes. 
	}
	\label{fig:honeycomb-k6-n72-all-wave-functions}
\end{figure}

We can see that each plaquette wave function has at least one local minimum.
While the exact position of the minima does not really matter, it is
interesting to note that for all wave functions but the $\zzz$ one, the minima
stay roughly stable. Indeed, we know that the $\zzz$ mean-field solution
consists of disconnected hexagons with $0$ flux. This solution is expected to
be captured by the $\zzz$ variational wave function, when $m$ is going to
infinity for a small value of $t_{d}/t_{h}$. This is indeed what can be
observed in the lower plot of \gref{honeycomb-k6-n72-all-wave-functions}: the
position of the $\zzz$ minimum moves to the left when $m$ increases but the
value of its energy remains higher than the energy of both the $\zpp$ and the
chiral wave function. By looking more carefully at the energies of the chiral
and $\zpp$ wave functions, it seems that the former becomes lower when $m$
increases and the latter remains stable. This behavior is the most
interesting feature of this analysis on a $72$-site cluster. Indeed, there is
a strong competition between the chiral and $\zpp$ wave functions when $m=1$
and for $m>1$ the energy of the chiral wave function becomes clearly lower.

\begin{figure}[h]
	\centering
	\includegraphics[scale=0.7]{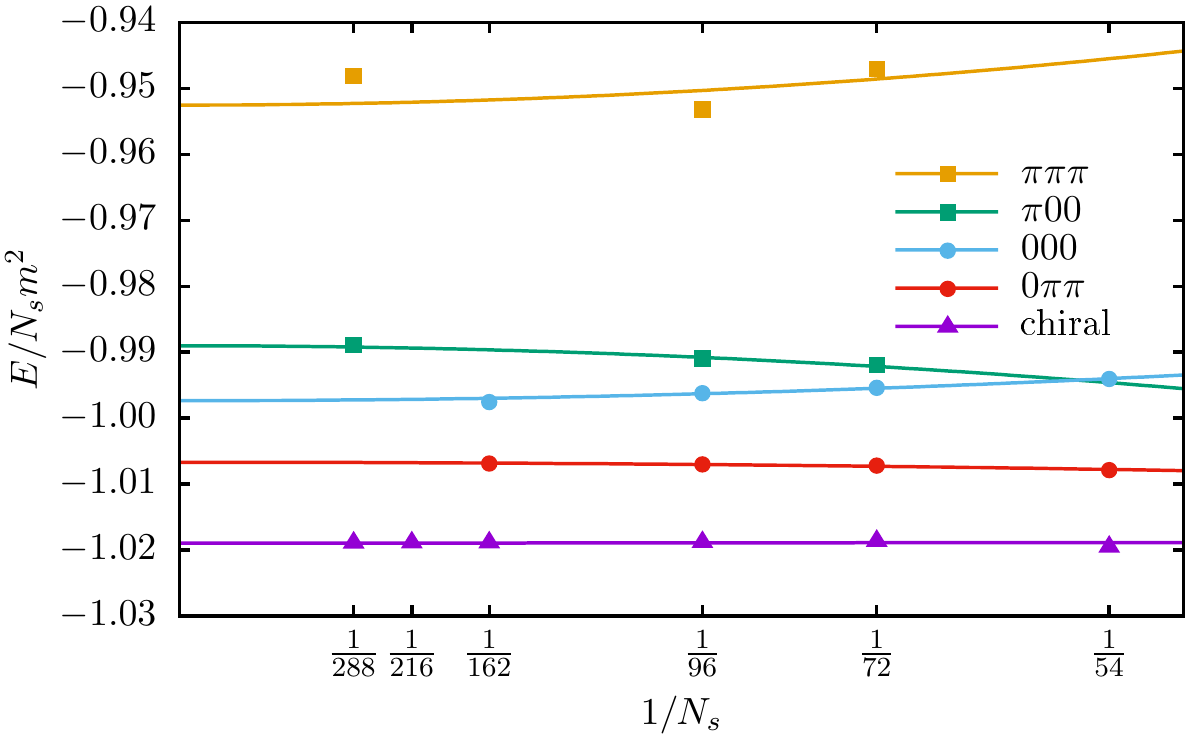}
	\caption{
	Finite size scaling for $m=3$. The function used for the fit is of the
	type $f(x)=ax^{2}+b$ for all data sets. For some system sizes, some values
	are missing. This is due to a degeneracy at the Fermi level. The error
	bars on the points are smaller than the symbols.
	}
	\label{fig:finite-size-scaling}
\end{figure}

The same study has been done for larger clusters (up to $288$ sites) and the
energies in the thermodynamic limit have been extrapolated. As an example,
\gref{finite-size-scaling} shows the variational energies as a function of the
system size for $m=3$. It is clear that the chiral wave function gives lower
energies than any of the plaquette ones no matter what the size of the system
is.

\begin{figure}[h]
	\centering
	\includegraphics[scale=0.7]{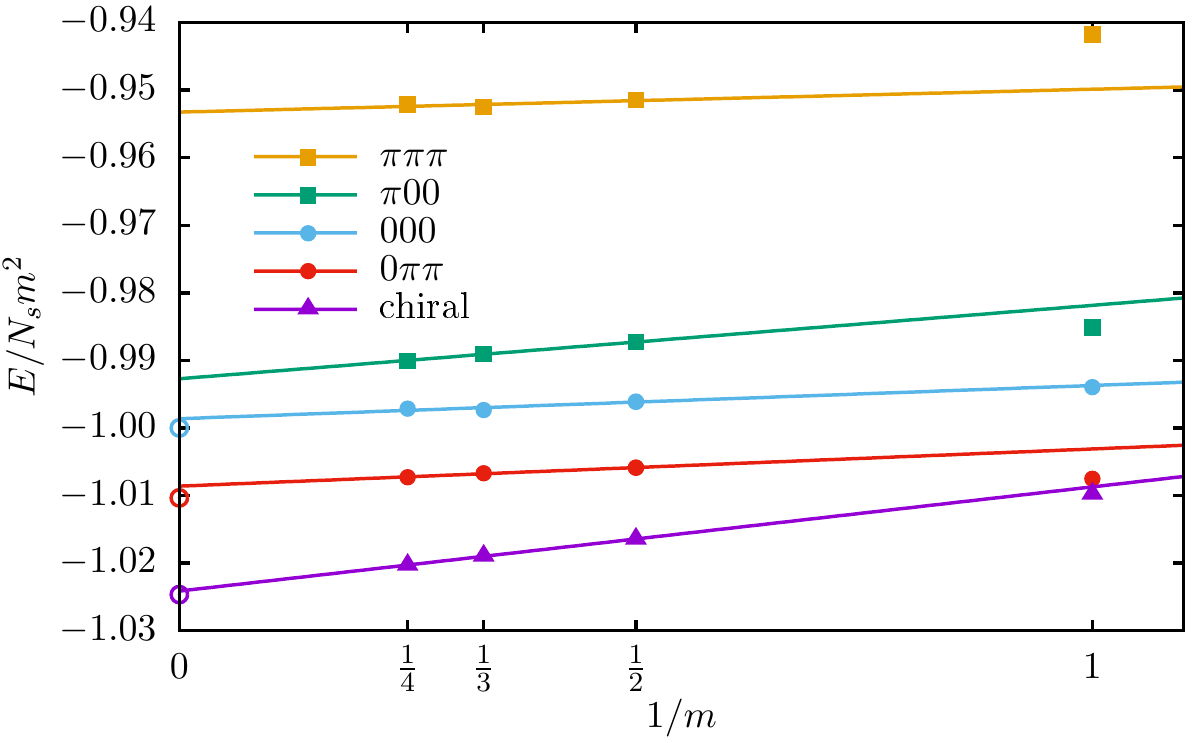}
	\caption{
	Energies per site obtained by VMC in the thermodynamic limit for different
	irreps. They are fitted by a line based on $m=2,3,4$ to show the agreement
	with the empty circles which are the mean-field results~\cite{Szirmai2011}
	valid for $m\rightarrow \infty$. 
	}
	\label{fig:honeycomb-k6-nthermo-energy-vs-m}
\end{figure}

The results shown in \gref{honeycomb-k6-nthermo-energy-vs-m} are the energies
extrapolated in the thermodynamic limit. Let us first focus on already
published results for the case with $m=1$. The chiral and $\zpp$ wave function
are in strong competition as already visible in
\gref{honeycomb-k6-n72-all-wave-functions}. It was numerically concluded on
the basis of extensive ED, VMC and iPEPS calculations~\cite{Nataf2016} that
the ground state is the $\zpp$ plaquette state. In the context of our
calculation, where we only calculate the energy of a single wave function, the
chiral state turns out to have a slightly lower energy, but as shown in
\citer{Nataf2016}, if several variational wave-functions with different
boundary conditions are coupled, this small energy difference is reverted in
favour of the plaquette phase. 

As soon as $m=2$, the energy of the chiral wave function becomes much lower
than the energy of any plaquette wave function, with an energy difference of
the same order of magnitude than that of the $m\rightarrow \infty$ case.
Moreover, this difference increases for larger values of $m$. The results for
$m>1$ can be fitted linearly in $1/m$, and the slope of the fit of the chiral
wave function energy is bigger than that of the plaquette wave function. Let
us note that the extrapolations of the fits to the limit $m=\infty$ of the
three lowest states (chiral, $\zpp$ and $\zzz$ plaquettes) agree with the
mean-field energies~\cite{Szirmai2011}, a good test of the validity of our VMC
simulations.

\section{Conclusion}
We have shown in the context of the \SU{N} model on the honeycomb lattice with
$N=6m$, where $m$ is the number of particles per site, that the presence of
chiral order in the mean-field limit ($m\rightarrow \infty$) is representative
of finite values of $m$ down to $m=2$. From that point of view, the case $m=1$
with its plaquette ground state appears as an exception. This is an
interesting step forward towards the stabilisation of chiral order in a simple
Heisenberg model with only nearest neighbour permutation and no ring-exchange
term. The first candidate in order of increasing $N$ is \SU{12} $m=2$. It is
still too large to be realized with alkaline rare earths, which are limited to
$N\leq 10$, but very close. This result suggests that a systematic
investigation of \SU{N} models with $N\leq 10$ for all compatible values of
$m$ (i.e. values of $m$ that divide $N$) and different lattice geometries
might indeed reveal a case of chiral order that could be stabilized with
alkaline rare earths and only nearest-neighbor permutations. Work is in
progress along these lines.

We acknowledge M. Lajko and P. Nataf for useful discussions, and K. Penc for a
critical reading of the manuscript. This work has been supported by the Swiss
National Science Foundation.

\bibliography{honeycomb-k6}
\end{document}